\documentclass[twocolumn,nofootinbib,amsmath,amssymb,aps,prd,balancelastpage,superscriptaddress]{revtex4-1}

\usepackage{color}
\usepackage[active]{srcltx}
\usepackage{amsmath,amsfonts,amssymb,amsthm,amstext,amscd,eucal,srcltx}
\usepackage{epsfig,graphicx,bm}
\usepackage{epstopdf, epsf}
\usepackage{dcolumn}
\usepackage{hyperref}

\newcommand{\be}{\begin{equation}}
\newcommand{\ee}{\end{equation}}

\newcommand{\bse}{\begin{subequations}}
\newcommand{\ese}{\end{subequations}}
\newcommand{\bea}{\begin{eqnarray}}
\newcommand{\eea}{\end{eqnarray}}
\newcommand{\ba}{\begin{array}}
\newcommand{\ea}{\end{array}}
\newcommand{\bc}{\begin{center}}
\newcommand{\ec}{\end{center}}

\begin{document}
\preprint{IPM/P-2012/009}  
\vspace*{3mm}

\title{Time-crystal ground state and production of gravitational waves\\ 
from QCD phase transition}%

\author{Andrea Addazi}
\email{andrea.addazi@lngs.infn.it}
\affiliation{Department of Physics \& Center for Field Theory and Particle Physics, Fudan University, 
200433 Shanghai, China}

\author{Antonino Marcian\`o}
\email{marciano@fudan.edu.cn}
\affiliation{Department of Physics \& Center for Field Theory and Particle Physics, 
Fudan University, 200433 Shanghai, China}

\author{Roman Pasechnik}
\email{Roman.Pasechnik@thep.lu.se}
\affiliation{Department of Astronomy and Theoretical Physics,
Lund University, SE-223 62 Lund, Sweden}
\affiliation{Departamento de F\'isica, CFM, Universidade Federal de Santa Catarina, C.P. 476, CEP 88.040-900, Florian\'opolis, SC, Brazil} \affiliation{Nuclear Physics Institute ASCR, 25068 \v{R}e\v{z}, Czech Republic}

\begin{abstract}
\noindent
We propose a novel mechanism for the production of gravitational waves in the early Universe that originates from the relaxation processes induced 
by the QCD phase transition. While the energy density of the quark-gluon mean-field is monotonously decaying in real time, its pressure undergoes 
a series of violent oscillations at the characteristic QCD time scales that generates a primordial multi-peaked gravitational waves signal in the radio 
frequencies' domain. The signal as an echo of the QCD phase transition, and is accessible by the FAST and SKA telescopes.
\end{abstract}

\maketitle

\section{Introduction}
\noindent 
The intriguing possibility that prompt phase transitions in the early Universe might have imprinted signatures in the background of gravitational radiation will be testable through the next generation of gravitational interferometers. The idea was firstly suggested in Refs.~\cite{Witten:1984rs,Turner:1990rc,Hogan:1986qda,Kosowsky:1991ua,Kamionkowski:1993fg}. 
New developments on the primordial gravitational waves (GW) production in the early Universe we achieved in Refs.~\cite{Hindmarsh:2013xza,Hindmarsh:2015qta}.
At the same time, recent studies on nuclear strong interaction provided several evidences for asymptotic freedom phenomena, including quarks confinement in baryons and mesons. Effects of confinement are related to the dimensional scale transmutation as much as first order phase transition (FOPT) phenomena, which are characterized by the dynamically generated energy scale $\Lambda_{\rm QCD}\simeq 200\, {\rm MeV}$ \cite{Aoki:2006we}. 
This suggested the possibility that the Quantum Chromodynamics (QCD) phase transition may generate a GW signal in the hot Early Universe, at a temperature of $T\simeq \Lambda_{\rm QCD}\simeq 200\, {\rm MeV}$. That a FOPT related to strong interactions may emit GWs was initially, although qualitatively, proposed by Witten \cite{Witten:1984rs}, and then quantitatively re-elaborated in Ref.\cite{Caprini:2010xv,Liu:2015psa}.    
The GW signal associated to the QCD phase transition (QCDPT) cannot be detected in GW terrestrial interferometers, such as LIGO/VIRGO \cite{LIGO} and KAGRA \cite{KAGRA}, cannot be either measured in 
future space experiments, such as LISA \cite{LISA}, U-DECIGO \cite{DECIGO}, BBO \cite{BBO}, TAIJI \cite{TAIJI} and TianQin \cite{TianQin} projects. The GWs frequency range of a QCDPT is around $10^{-8}\div 10^{-9}\, {\rm Hz}$, which is 5-6th digits lower than the one provided by space experiments, and 9-10th digit far from LIGO/VIRGO/KAGRA \cite{Liu:2015psa}. Furthermore, QCDPT does not leave any smoking-gun imprinting in the Cosmic Microwave Background, which is sensitive to very low frequency modes (5th-6th digits less \cite{Liu:2015psa}). 

Nonetheless, a nHz phase transition such as a QCDPT can be detected, with high precision, from radio astronomical observation 
of pulsar timing: the GW backgrounds propagating through pulsar systems alter the radio signal, leaving an imprinting that is principle observable. This opens a pathway towards the exciting possibility of testing fundamental particle physics with current and future radio astronomy experiments, including FAST \cite{FASTweb}, and SKA \cite{SKAweb}. Within previous QCDPT analyses, the role of possible relaxation phenomena and gravitational back-reactions were completely neglected. But after the QCDPT, prompt and violent relaxation effects around the QCD vacuum energy state are expected, which retains a broad analogy with the reheating mechanism in inflationary models. 

In this letter, we study in detail the possible effects of the gluon condensate relaxation phenomena. We analyze the non-linear field equations for the gluonic condensate, coupled to the Einstein equation, in a Friedmann-Lema\^itre-Robertson-Walker (FLRW) cosmological background. During the relaxation phase, a surprising non-equilibrium phenomenon arises: the gluonic condensate field violently oscillates during the relaxation phase, inducing fast oscillations of the energy-momentum tensor trace for a transient time of $\tau \simeq 10\div 20 \Lambda_{\rm QCD}^{-1}$. The oscillating solution is a classical non-perturbative solution of the Yang-Mills field equations coupled to the Einstein field equations. The emergence of spikes, localized in a characteristic QCD time lapse $\Delta t\simeq \Lambda_{\rm QCD}^{-1}$,
and extended in the space dimensions, reveals the presence of a ordered pattern of space-like soliton/domain-walls solutions. We dub these new solutions {\it chronons}. After a cosmological time $t > 20 \Lambda_{\rm QCD}^{-1}$, the spikes' periodicity disappears, and the energy density approaches the QCD vacuum energy minimum. The time-ordered classical solution that we found is a {\it time crystal}, i.e. a periodic classical solution spontaneously breaking time invariance down to a discrete time shift symmetry $T_{n}:t \rightarrow t+n \Lambda_{\rm QCD}^{-1}$, $n$ denoting a natural number. The concept of time crystal has been first proposed by Wilczek in Refs.~\cite{Wilczek:2012jt,Wilczek-2} within the context of superconductors and superfluids physics\footnote{The original implementation of this idea was criticized in Refs.~\cite{Bruno:2013rdc,Watanabe:2014hea}.}. For a review of time crystals, see e.g. Ref.~\cite{Sacha:2017fqe}. The experimental discovery of time crystals was achieved in Refs.~\cite{Autti:2017jcw}.
 The spontaneous symmetry breaking of $T$-$invariance$  from the localization of chronons is associated to the appearance of Nambu-Goldstone bosons, as time-like moduli excitations over the classical background. 
 
During the relaxation stage, a new characteristic feature in the produced GW signal. While the energy-density part of the energy momentum tensor does not exhibit so violent transitions, the condensate pressure provides the main contribution to the energy-momentum tensor trace variation. These pressure kinks inject kinetic energy into the primordial plasma, inducing turbulence and sound/shock waves in the plasma very efficiently. In analogy with the case of bubble propagating in the plasma, the gravitational radiation is emitted from magnetohydrodynamical (MHD) turbulence and sound waves. From our numerical simulations, which we compare with semi-analytical estimates, we show that such gravitational background signal can be tested in future radio observatories form pulsar timing effects. The spectrum that is predicted not only lies within the SKA sensitivity, but it further displays very peculiar features of the shape form that cannot be reproduced in any other known mechanism. In other words, time crystallization of QCD during the relaxation phase can be tested in next future, which implies a radical reconsideration of our picture of QCD confinement itself from the prospective of dynamical cosmological evolution.

\section{Space-like domain walls from T-breaking}
\noindent
A standard static domain-wall can be easily obtained from a scalar field theory that is $Z_{2}$ invariant. With a simple sombrero-like Higgs potential, $Z_{2}$ can be spontaneously broken when the scalar field rolls down to one of the two possible minima $\phi_{vac}=\pm v$. These internal field configurations can be localized in the space direction $z$ as kink profiles. The kink profile interpolates the two minima, namely $\phi(z=-\infty)=-v$ and $\phi(z=\infty)=v$. A domain-wall configuration, as a $xy$-plane orthogonal to the z-direction, is achieved through the kink profile transition region, and its characteristic thick in z-direction is directly related to the kink shape. For a $\lambda \phi^{4}$ theory with sombrero potential, one can find a simple analytic kink solution, specified by $\phi(z)=v\tanh\Big[\frac{\lambda v}{\sqrt{2}}(z-z_{0})\Big]$, with $z_{0}$ the kink center.  

As well known, for standard domain-walls the translational invariance is spontaneously broken, being the barrier localized in a $z_{0}$ point. This corresponds to the appearance of a Nambu-Goldstone modulus boson $z_{0}(t,x,y)$, localized on the surface of the domain wall, as a low energy-excitation of its surface in the z-direction. 

Intriguingly and exotically, one may consider a kink profile that, despite of been localized in a space direction, it is localized in time. A new domain wall extended in three spatial dimension but localized in a time lapse, which we dub chronons, may correspond to this solution. By just replacing the z-coordinate with the time variable, one can consider a kink solution, such as $\phi(t)=v\tanh\Big[\frac{\lambda v}{\sqrt{2}}(t-t_{0})\Big]$, centered in a time instant $t=t_{0}$ and interpolating the two vacuum states in the asymptotic time limits $\phi(t=-\infty)=-v$ and $\phi(t=\infty)=v$. This solution is associated to a spontaneous symmetry breaking of the time invariance and to the appearance of a Nambu-Goldstone boson localized on the xyz surface $t_{0}(x,y,z)$. 
 
In the case of the gluon condensate field equation coupled to gravity, in a FLRW cosmological background one can decompose the gluonic field in a classical background field $U(t)$ plus a non-homogeneous part --- see the Appendix for more technical details. Let us consider the limit of a static FLRW space time ($a={\rm cost}$). A branch of solutions for the $U$ field satisfies the equation 
\begin{equation}
    \label{eed}
  U'^{2}-\frac{1}{4}U^{4}={\rm const}\,,
\end{equation}
 where $U'$ is the field derivative with respect to the Cartesian coordinate time, which we denote here with $x_0\equiv \eta$. A branch of solutions of these equations, obtained  by $U^{2}\rightarrow U^{2}-U_{0}^{2}$ energy density vacuum shift, corresponds to kink (antikink) profiles 
 \begin{equation}
 \label{UUU}
 U(\eta)\simeq  \frac{v}{\sqrt{2}}\tanh[\frac{v}{\sqrt{2}}(\eta-\eta_{0})]\, . 
 \end{equation}
 where $v\simeq \Lambda_{\rm QCD}$. A space-like domain wall corresponds to a kink profile of this type. Time-translation is spontaneously broken, and a $\eta_{0}(x,y,z)$ moduli field arises, with $U$ acquiring the dependence
$U(\eta-\eta_{0}(x,y,z))$. The coordinate $x,y,z$ are the domain-walls worldsheet coordinates. The effective corresponding action reads as
\begin{eqnarray}
    \label{Acef}
 S=&&\int\! d^{4}x \,\frac{1}{2}\Big[\Big(\frac{\partial \phi}{\partial \eta_{0}}\frac{\partial \eta_{0}}{\partial x^{a}}\Big)^{2}-V(U)  \Big] \nonumber\\
 =&& \, \, {\rm const}+\frac{T_{W}}{2}\int \!d^{3}x\, \Big(\frac{\partial \eta_{0}(x^{a})}{\partial x^{a}}\Big)^{2}\, .  
 \end{eqnarray}
This shows that the moduli field is massless, according to the Nambu-Goldstone theorem.

When the gravitational dynamics is taken into account, and the scale factor time-dependence is considered, a more complicated time pattern for the space-like domain walls arises --- see the Appendix for all the technical details. In this latter case, time-translation is not only broken down to a  $Z_{2}$ symmetry involving the $T_{+}$ and $T_{-}$ branches, but more interestingly is spontaneously broken down to a discrete time-translation 
symmetry $T_{n}:\eta\rightarrow \eta+n\Lambda_{\rm QCD}^{-1}$. In other words, the system behaves as a time crystal.

\begin{figure}[t]
\centerline{ \includegraphics [height=5.cm,width=0.9\columnwidth]{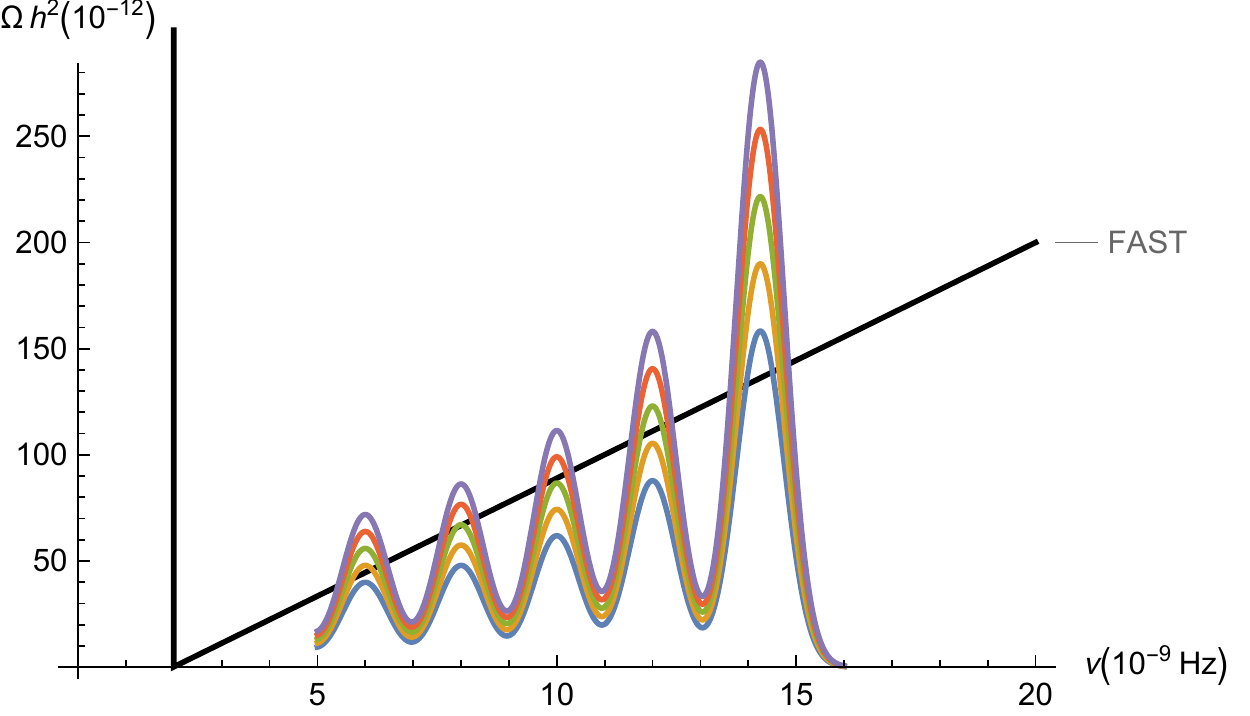}}
\caption{The gravitational waves spectrum is displayed for different efficiency factors, in comparison with FAST sensitivity curve \cite{FASTweb}. The efficiency factor considered are $\kappa=0.03\div 0.1$.  }
\label{f1}
\end{figure}

\begin{figure}[t]
\centerline{ \includegraphics [height=5.cm,width=0.9\columnwidth]{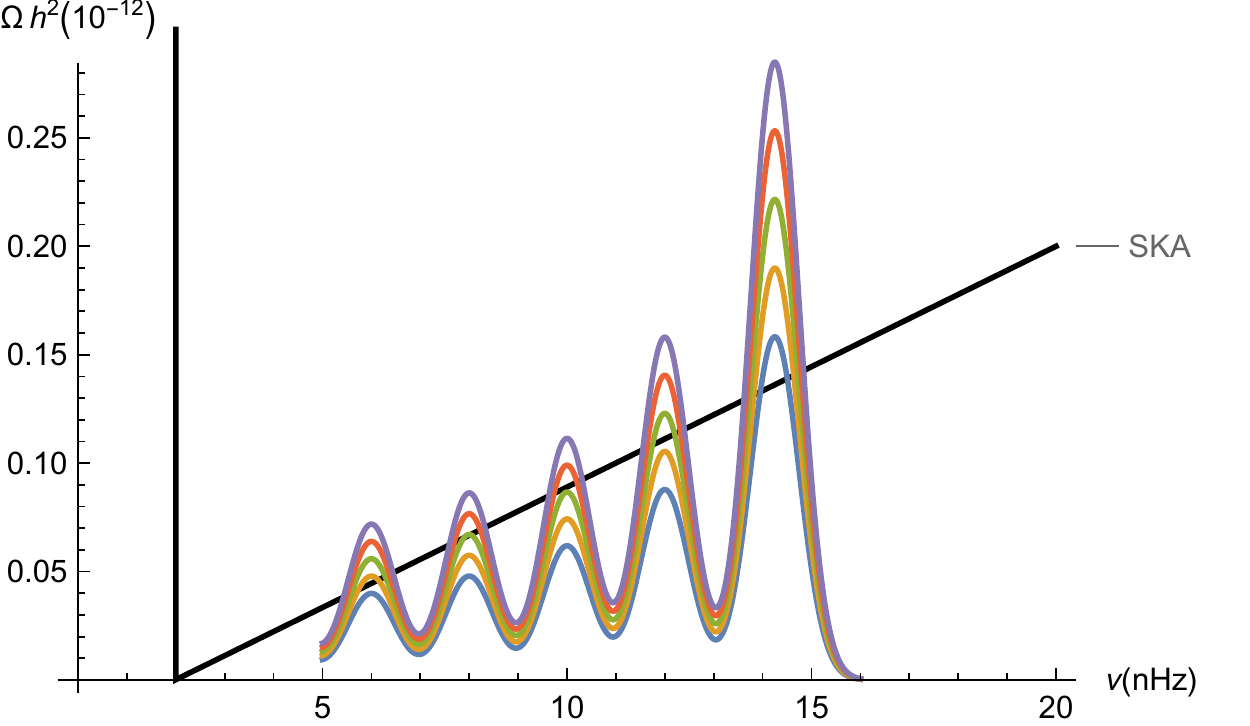}}
\caption{The gravitational waves spectrum is displayed for different efficiency factors, in comparison with FAST sensitivity curve \cite{SKAweb}. The efficiency factor considered range in $\kappa=10^{-3}\div 3\times 10^{-3}$.}
\label{f2}
\end{figure}

\section{Gravitational wave emission}
\noindent 
The general coupled field equations of gluon field with gravity reads
\begin{eqnarray}
&&\frac{1}{\varkappa}\left(R_\mu^\nu-\frac12\delta_\mu^\nu R\right)=
\frac{b}{32\pi^2}\frac{1}{\sqrt{-g}}\biggl[\biggl(-\mathcal{F}_{\mu\lambda}^a\mathcal{F}^{\nu\lambda}_a
\nonumber
\\
&& \quad +\,\frac14\delta_\mu^\nu
\mathcal{F}_{\sigma\lambda}^a\mathcal{F}^{\sigma\lambda}_a\biggr)
\ln\frac{e|\mathcal{F}_{\alpha\beta}^a\mathcal{F}^{\alpha\beta}_a|}{\sqrt{-g}\,
\lambda^4}-\frac14 \delta_\mu^\nu \,
\mathcal{F}_{\sigma\lambda}^a\mathcal{F}^{\sigma\lambda}_a\biggr]\,, \\
&& \left(\frac{\delta^{ab}}{\sqrt{-g}}\partial_\nu\sqrt{-g}-f^{abc}\mathcal{A}_\nu^c\right)
\left(\frac{\mathcal{F}_b^{\mu\nu}}{\sqrt{-g}}\,\ln\frac{e|\mathcal{F}_{\alpha\beta}^a
\mathcal{F}^{\alpha\beta}_a|}{\sqrt{-g}\,\lambda^4}\right)=0\,, \nonumber
\end{eqnarray}
where $\lambda\equiv \xi \Lambda_{\rm QCD}$ is related to the QCD scale parameter $\Lambda_{\rm QCD}$ by an arbitrary scaling constant $\xi$ and $e$ is the base of the natural logarithm. In the FLRW background, here cast in conformal coordinates defined by $ds^2=a^2(\eta)(d\eta^2 + d\vec{x}^2)$, the dynamical system is simplified to 
\begin{eqnarray}
&& \frac{6}{\varkappa} \frac{a''}{a^3} = T^{\mu,{\rm U}}_{\mu}\,, \nonumber \\
&& T^{\mu,{\rm U}}_{\mu}=\frac{3b}{16\pi^2 a^4}\Big[(U')^2-\frac{1}{4}U^4\Big]\,, \nonumber \\
&& \frac{\partial}{\partial \eta}\Big(U'\,\ln\frac{6e\big|(U')^2-\frac{1}{4}U^4\big|}{a^4\lambda^4}\Big) \nonumber \\ 
&& \qquad +\,\frac{1}{2}U^3\,\ln\frac{6e\big|(U')^2-\frac{1}{4}U^4\big|}{a^4\lambda^4}=0 \,. 
\end{eqnarray}

In order to account for thermal bath effects, we have to consider the thermal loop correction to the classical equations. The leading order corrections are proportional to $T^{2}U^{2}$, where $T$ is the temperature of the early Universe plasma. 

Solving the dynamical system of equations specified above, at $T=0$, we find a $U(\eta)$ profile characterized by a time-ordered pattern of spikes --- see the Appendix for more technical details. Turning on thermal corrections, the only relevant $O(1)$ corrections arise in correspondence of the first spike, close to the QCD phase transition scale $\Lambda_{\rm QCD}$. After $10\,\Lambda_{\rm QCD}^{-1}$, thermal correction will reduce to a $1\%$ order. The trace of energy-momentum tensor follows the spike series solution, with violent oscillations while having relaxation. Relaxation is induced by the linear dissipative terms of the gluonic condensate in the Field Equations, while spikes can be understood as a back-reaction effect of the gluonic condensate with the gravitational background. The energy-density part $\rho_{U}$ has only a suppressed modulation over the relaxation decay profile in time. Since the gravitational waves emission is related to the time variation of the energy-density, it turns out that the GW spectrum that can be derived is actually suppressed. Most of the trace tensor variation are provided by the pressure component $p_{U}$. The pressure kinks can be efficiently transmitted to the primordial plasma, since the gluonic condensate is strongly interacting with it. Consequently, the pressure kinks pattern chaotized the early Universe plasma, inducing turbulence and shock sound waves in it. As it happens in the standard picture of first order phase transitions, turbulence wiggles and sound waves efficiently produce gravitational radiation. The pressure distribution of one kink in the plasma thermalizes very efficiently, within a standard deviation of the order of the QCD scale. Numerical results are shown in Fig.~\ref{f1} and Fig.~\ref{f2}, and put in comparison with current FAST experiment sensitivity curves and future SKA predicted bounds. The model efficiency factors enter crucially the pressure transfer mechanism from the classical condensate to the Universe plasma. Although clearly affected by several uncertainties, one can show several GW profiles in the reasonable efficiency range of $10^{-2}\div 10^{-3}$. We remark that an efficiency lower than $10^{-3}$ seems to be nearly impossible, since the gluonic condensate cannot be so weakly coupled to the plasma. This observation carries important phenomenological consequences. Indeed, we can show that, for an efficiency higher than the $3\%$, FAST will definitely probe this model. On the other hand, future FAST data, one can reach the $0.1\%$ efficiency scale. 

We provide below simple semi-analytic estimates, which are nonetheless in agreement with our numerical analysis. 

The red-shift due to the expansion of the gravitational background must be taken into account while comparing the GW signals {\it ab origine} with measurements at present time. At this purpose, we recall that the ratio between the scale factor of the Universe today $R_{0}$ and the scale factor of the Universe during the GW production is expressed by
\begin{equation}
\label{RsRz}
\frac{R_{*}}{R_{0}}=8.0\times 10^{-14}\left(\frac{100}{g_{*}}\right)^{\frac{1}{3}}\left(\frac{1\, {\rm GeV}}{T_{*}}\right)\, , 
\end{equation}
$g_*$ denoting the effective number of degrees of freedom. Assuming that the Universe expanded adiabatically implies that the entropy $S\sim R^{3}\, T^{3}$ remained constant. The characteristic frequency of the GW signal today, denoted as $f_{0}$, is related to the one on the GW emission time $f_{*}$
by
\begin{equation}
\label{frequency}
f_{0}=f_{*}\left(\frac{R_{*}}{R_{0}}\right)=1.65 \times 10^{-7}{\rm Hz}\! \left(\frac{f_{*}}{H_{*}}\right)\! \left(\frac{T_{*}}{1\, {\rm GeV}}\right)\left(\frac{g_{*}}{100}\right)^{\frac{1}{6}}\!\! . 
\end{equation}
The order of magnitude of the GW energy density today, denoted as $\Omega_{\rm GW}$, is related to the one during the emission time, namely $\Omega_{\rm GW *}$, through 
\begin{eqnarray}
\label{OmegaGW}
\Omega_{\rm GW}&& =\Omega_{\rm GW*}\left(\frac{R_{*}}{R_{0}}\right)^{4}\left(\frac{H_{*}}{H_{0}}\right)^{2} \nonumber \\ 
&&=1.67\times 10^{-5}h^{-2}\left(\frac{100}{g_{*}}\right)^{\frac{1}{3}}\Omega_{\rm GW*}\, ,
\end{eqnarray}
where $h$ is the current value of the Hubble parameter in units of 
$100\, {\rm km}\,/ ({\rm sec}\, {\rm Mpc})$, and 
\begin{equation}
\label{Hs}
H_{*}=\frac{8\pi G \rho_{\rm rad}}{3}=\frac{8\pi^{3}g_{*}T_{*}^{4}}{90M_{Pl}^{2}}\, , 
\end{equation}
which is the Hubble contribution in the radiation dominated epoch. 

The sound and turbulence spectrum induced by the spiky pressure kinks is in general very complicated. However, it will display a characteristic series of peaks, related to the pressure peaks. The magnitude of these GW peaks can be estimated very easily thanks to semi-analytical estimates. The turbulence GW peaks are described by
\begin{equation}
\label{hajk}
h^{2}\Omega_{\rm turb}=3.35\times 10^{-4}\sum_{i}^{N_{\rm eff}}\!\left(\frac{H_{*,i}}{\beta_{i}}\right)\!\left(\frac{\kappa_{turb}\alpha}{1+\alpha} \right)^{\frac{3}{2}}\!\left(\frac{100}{g_{*,i}}\right)^{\frac{1}{3}}\!\!v,
\end{equation}
\begin{equation}
\label{fgak}
f_{\rm turb}=2.7\times 10^{-2}\, {\rm mHz} \frac{1}{v}\left(\frac{\beta_{i}}{H_{*,i}}\right)\left(\frac{E_{*,i}}{{\rm GeV}}\right)\left( \frac{g_{*,i}}{100}\right)^{\frac{1}{6}},
\end{equation}
where the sum is over the number of peaks that contribute significantly to the GW spectrum, and 
\begin{equation}
\label{akak}
\alpha=\frac{\rho_{U}}{\rho_{\rm rad}}
\end{equation}
is the ratio between the energy density $U$ and the radiation energy density. 

The sound waves spectrum is characterized by the expressions 
\begin{equation}
\label{gak}
h^{2}\Omega_{\rm sound}=2.65\times 10^{-6}\sum_{i=1}^{N_{\rm eff}}\!\left( \frac{H_{*,i}}{\beta_{i}}\right)\!\left(\frac{\kappa_{v}\alpha}{1+\alpha}\right)^{2}\!\! \left(\frac{100}{g_{*,i}} \right)^{\frac{1}{3}}\!\!v_{i},
\end{equation}
\begin{equation}
\label{hjak}
f_{\rm sw}=1.9\times 10^{-2} {\rm mHz}\! \sum_{i=1}^{N_{\rm eff}}\frac{1}{v_{i}}\left(\frac{\beta_{i}}{H_{*,i}}\right)\!\! \left(\frac{T_{*,i}}{100\, {\rm GeV}}\right)\!\!\left(\frac{g_{*,i}}{100}\right)^{\frac{1}{6}}. 
\end{equation}

We estimated the $U$ peak rapidity as $v_{i}=\Delta (T_{\mu}^{\mu}) \beta_{i}$, which is close to $1$ (fast), and the inverse time scale of the peaks as $\beta_{i} \simeq \Lambda$ ($\Lambda$ being the confinement scale), while $\kappa_{\rm v, turb}$ are the efficiency factors of the energy transfer from the condensate peaks to the early Universe plasma. We find that: i) the energy-scale of the first peaks is around $200\, {\rm MeV}$; ii)  $\beta_{i} \simeq H_{*,i}$, since $H_{*}$ is related to the energy density and pressure of the Universe and follows the oscillations of the condensate; iii) $\alpha \simeq 1$, if the condensate dominates in the radiation epoch. 

Assuming the efficiency factors $\kappa_{\rm turb,v}\simeq 0.1\%\div 1\%$ --- in analogy to typical response factors in the plasma during FOPTs ---  from the estimates specified above, we obtain an energy frequency within the frequency range $10^{-9}\, \div 10^{-8}\, {\rm Hz}$ for both turbulence and sound waves. This implies that the GW signal here predicted lies in the radio-astronomy pulsar timing scale, while the energy-density of the GW signal for both the contributions is around $10^{-8}\div 10^{-12}$. Thus, SKA must be definitely able probe to this GW signal. 

\section{Conclusions}
\noindent 
We have shown that the relaxation dynamics of the gluon condensate close to the QCD phase transition behaves as a time-crystal within a time range of $1\div 20\Lambda_{QCD^{-1}}$. This is an effect obtained as a non-equilibrium solution from the classical gluon equations of the gauge fields coupled with gravity. Time translation is spontaneously broken in a time translation discrete symmetry. We have shown that this proposal to model gluonic condensates in the early Universe cosmology is already testable. The model we developed predicts a gravitational radiation background that can alter the pulsar timing system, and thus can be tested in radio-astronomy experiments. Specifically, predictions of the model lay in sensitivity curves of FAST and SKA. A clear understanding of confinement remains the most challenging problem of the Standard Model of particles physics. Possible informations about the confinement dynamics in the early Universe from radio-astronomy opens a new pathway towards the frontiers of strong interaction physics. 

\appendix

\section{Spatially homogeneous isotropic YM condensates}
\label{Sec:YM-condensates}
\noindent 
We discuss in this section how a gauge-invariant description of spatially homogeneous isotropic Yang-Mills (YM) condensates that only depend only on time can be obtained. For this purpose, it is most useful to work in the ghost-free temporal (Hamilton or Weyl) gauge, fixed by the conditions
\begin{equation}
A^a_0=0\,, \label{gauge}
\end{equation}
which is the basis of the Hamiltonian formulation. In this gauge, the asymptotic states of the $S$-matrix automatically contain transverse modes only, which enables to formulate the YM theory in the Heisenberg representation, consistently beyond the phase transitions --- for more details, see e.g. Ref.~\cite{Bogolubov,Faddeev}. Under the condition (\ref{gauge}), the Gauss-laws, which enforce the symmetry under the residual time-independent gauge 
transformations, have to be implemented additionally to the Hamiltonian equations of motion. These equations can be brought into an unconstrained form by imposing a further time-independent gauge. While in the perturbative phase the non-Abelian Coulomb gauge is a useful choice, in the case of spatially homogeneous condensates the symmetric gauge for SU(2) \cite{Pasechnik:2013sga,KP1,KMPR,Prokhorov:2013xba,Dona:2015xia} and $SU(3)$ \cite{Pavel2012,Pavel2013,Pavel2014,Addazi:2016sot,Alexander:2016xbm,Addazi:2016nok} gauge theories is a suitable choice as will be discussed below.

\subsection{Gauge invariant description of a coherent SU(2) condensate}
\label{Sec:cond-SU2}
\noindent 
In the SU(2) gauge theory, due to the local isomorphism of the isotopic SU(2) gauge group and the SO(3) group of spatial 3-rotations, the unique (up to a rescaling) SU(2) YM configuration can be parameterized in terms of a scalar time-dependent spatially-homogeneous field (see e.g. Refs.~\cite{KP1,KMPR,Cervero:1978db,Henneaux:1982vs,Hosotani:1984wj}). Indeed, in the case of the SU(2) Yang-Mills fields 
\begin{equation}
A_k=A_{a k} {\sigma_a\over 2}\,, \qquad 
\Big[{\sigma_a\over 2},\,{\sigma_b\over 2}\Big] = 
i \epsilon_{abc} {\sigma_c\over 2}\,, 
\end{equation}
one employs the polar decomposition (symmetric gauge) \cite{KP1,KMPR} of the gauge field
\begin{equation}
\label{coordtrafo}
A_{ak}(q, S) = O_{ai}(q)\, S_{ik} - {1\over 2\bar{g}}\epsilon_{abc}\, 
\big(O(q)\partial_k O^T(q)\big)_{bc} \,,
\end{equation}
into an orthogonal matrix $O$, depending on three gauge angles, and a physical symmetric (positively definite) tensor field $S_{ik}$ ($\epsilon_{ijk}\,S_{jk}=0$) with two spatial indices consisting of one spin-0 component $S_{ik}^{(0)}=(1/3)\delta_{ik}\,\mathrm{Tr}\,S$ and five spin-2 components. Using the symmetric gauge, one therefore obtains a unique and gauge-invariant decomposition of the gauge field into a spatially homogeneous isotropic part (the YM condensate) and non-isotropic/non-homogeneous parts (the YM waves), namely 
\begin{eqnarray} \label{SSU2}
&& A_{ak}\big(t,\vec{x}\big) = \delta_{ak} U(t) + \widetilde{A}_{ak}\big(t,\vec{x}\big)\,, \\ 
&& \langle \widetilde{A}_{ik}\big(t,\vec x\big) \rangle = \int d^3x\; \widetilde{A}_{ik}\big(t,\vec x\big) = 0 \,,
\nonumber
\end{eqnarray}
where the YM condensate is positively definite $U(t)>0$. In the QFT formulation, the inhomogeneous YM wave modes $\widetilde{A}_{ik}$ are interpreted as YM quanta (e.g. gluons), while $U(t)$ contributes to the ground state of the theory. This is consistent with taking the expectation value of the fields on a coherent state $|\alpha\rangle$ that encodes the symmetry of the background and represents a long-wavelength cosmological condensate state $\langle \alpha| A_{ak} | \alpha \rangle=\delta_{ak} U(t)$, having picked at $\vec k =0$ the macroscopic semiclassical-state.

In the absence of gravity, the spatially homogeneous isotropic part  $\delta_{ik}U(t)$ of 
Eq.~(\ref{SSU2}) satisfies the classical YM equations 
\begin{equation}
 (\dot{U})^2 + \bar{g}^2 \, U^4 = \mathrm{const} \,, \label{USU2}
\end{equation}
which can be integrated analytically \cite{Pasechnik:2013sga}. In the FLRW Universe, the classical YM condensate behaves as radiation medium with $a(t)\propto t^{1/2}$ and $p_{\rm YM}=\epsilon_{\rm vac}/3$, characteristic features of the classical YM field behaviour. The semi-classical dynamics of the homogeneous SU(2) condensate and the small YM waves has been thoroughly studied in the Minkowski spacetime in Ref.~\cite{Prokhorov:2013xba}. 

From the group-theoretical viewpoint, the separation into spatially homogeneous and inhomogeneous components (\ref{SSU2}) in the Minkowski spacetime has certain similarities with an analogical procedure in the conventional QCD instanton theory in Euclidean spacetime \cite{Boucaud:2002nc,Hutter:1995sc,Schafer:1996wv}. Namely, in both cases one performs a mapping of 3-space onto SU(2) subgroup elements of the color SU(3)$_c$. Moreover, as will be discussed below, both cases exhibit remarkable dynamical similarities (or duality). 

Notice furthermore that the homogeneous YM condensate can be introduced for every gauge group, which contains at least one SU(2) subgroup (e.g. SU(N)), whereas the condensate embedding or the extraction procedure can be different in every case. These  condensates obey similar equations of motion which may differ by a rescaling of the coupling constant. Hence, the frequency of the classical YM condensate oscillations for a SU(N) gauge theory depends on integer $N$, while the overall dynamical properties are the same. 

\subsection{Inclusion of matter fields: effective QCD energy-momentum tensor}
\label{Sec:matter}
\noindent 
The effective QCD energy-momentum tensor that includes also quarks is expressed by relations that are similar to ones derived for the case of pure gluodynamics. The only difference between the two expressions arises in the effective beta-function coefficient $b$.
When only gluons are taken into account, the one-loop $\beta$-function coefficient of the pure gluodynamics provides the value $b=b(0)=11$. When quarks are also included, the effective value is found to be \cite{Pasechnik:2013poa}
\begin{eqnarray} \label{beff} 
b_{\rm eff}=b +8 L_g (m_u + m_d +m_s)\simeq 9.6 \,,
\end{eqnarray}  
in which $L_{g}=(1500\pm 300\, {\rm MeV})^{-1}$ is the correlation length of the fluctuations, recovered by the experimental data on quark and gluon condensates (and supported by the lattice QCD calculations), and with a value  close to the minimal scale of quantum-topological fluctuations that contribute to the QCD vacuum. It is indeed renown that for quantum-topological quark-gluon fluctuations \cite{Schafer:1996wv} the equality holds 
\begin{eqnarray}
&&\langle 0|  :\bar{s} s : |0 \rangle \simeq \langle 0|  :\bar{u} u : |0 \rangle \langle 0|  :\bar{u} u : |0 \rangle \label{tqgf}\\
&&= - \langle 0|  : \frac{\alpha_s}{\pi} F^a_{\mu \nu} F_a^{\mu \nu}: |0 \rangle L_g = - (225 \pm 25\, {\rm MeV})^3\,.  \nonumber
\end{eqnarray}  
From a phenomenological point of view, we may deploy reasonable assumptions, and imagine non-perturbative quantum-wave (hadron) fluctuations to occur at the same space-time scales as quantum topological fluctuations. This implies that they should satisfy a functional relation in analogy to \eqref{tqgf}. The operator relation between quark and gluon fluctuations can be then established in terms of the trace of the quark energy-momentum tensor, once the vacuum average $\frac{1}{4} \langle 0 | T^\mu_{\ \mu \,{\rm (QCD)}}  | 0\rangle $ is performed, and thus the topological contribution to the energy density recovered. The trace is hence derived from the trace anomaly \cite{T1,T2,T3}, namely 
\begin{eqnarray}
T^\mu_{\ \mu \,{\rm (QCD)}} = \frac{\beta(g_s^2)}{2} \, F^a_{\mu \nu} F_a^{\mu \nu} + \sum \limits_{q=u,d,s} m_q \bar{q} q\,.
\end{eqnarray}
The characteristic topological instanton-type contribution to the energy density of the QCD vacuum, here denoted with $\epsilon^{\rm QCD}_{\rm top.}$, is then recovered to be
\begin{eqnarray}
\epsilon^{\rm QCD}_{\rm top.}\!&=&\! -\frac{9}{32}  \langle 0 | : \frac{\alpha_s}{\pi} \, F^a_{\mu \nu} F_a^{\mu \nu}  : | 0\rangle + \frac{1}{4} \Big[  \langle 0 | : m_u \bar{u} u : | 0\rangle  \nonumber \\
\!\!&\phantom{a}&\!\! +\langle 0 | :  m_d \bar{d} d : | 0\rangle   + \langle 0 | :  m_s \bar{s} s : | 0\rangle \Big]  \nonumber \\
\!\!&\phantom{a}&\!\!  \simeq - (5 \pm1)\, \times 10^9 \, {\rm MeV}\,. \label{eps}
\end{eqnarray}
This expression is due to the gluons and the light sea $u$, $d$, $s$ quark contributions, providing the maximal value of the topological contribution to the QCD vacuum energy density. Nonetheless, thanks to our assumptions both the relations \eqref{tqgf} and \eqref{eps} are also valid for the quantum-wave contributions, hence providing the effective quark contribution to the QCD energy momentum tensor operators, i.e.
\begin{eqnarray}
T^\mu_{\ \nu \, {\rm (q),\, eff }} = \frac{8 L_g}{b(3)} (m_u+m_d+m_s) \, T^\mu_{\ \nu \, {\rm (g),\, eff }} \,,
\end{eqnarray}
with obvious meaning of the labels. From this latter, and considering the expression of the operator energy-momentum tensor of the gluon field
\begin{eqnarray}
T^\mu_{\ \nu \, {\rm (q),\, eff }} = \frac{8 L_g}{b(3)} (m_u+m_d+m_s) \, T^\mu_{\ \nu \, {\rm (g),\, eff }} \,,
\end{eqnarray}
one can recover in the one-loop approximation the phenomenologically motivated complete QCD energy-momentum tensor
\begin{eqnarray}
T^\mu_{\ \nu \, {\rm (QCD) }} \! & \simeq &  \! \frac{b_{\rm eff}}{32 \pi^2} \left( -
F^a_{\nu\rho}\, F_a^{\mu\rho} + \frac{1}{4} \delta^\mu_\nu \, F^a_{\mu\nu}\, F_a^{\mu\nu} \right) \ln \frac{e \mathcal{J}}{\lambda^4} \nonumber\\
\!\!&\phantom{a}&\!\! -\delta^\mu_\nu \, \frac{b_{\rm eff}}{128 \pi^2}\, F^a_{\mu\nu}\, F_a^{\mu\nu}\,,
\end{eqnarray}
with $\mathcal{J}=-F^a_{\mu\nu}\, F_a^{\mu\nu}/\sqrt{-g}$, the constant $\lambda$ related to the QCD scale parameter by a parameter $\xi$, namely $\lambda\equiv \xi \Lambda_{\rm QCD}$, $e$  denoting the base of the natural logarithm and $b_{\rm eff}$ specified in \eqref{beff}.

\section{Homogeneous gluon condensate evolution}
\label{Sec:cond-PT}
\noindent
Now we come to an analysis of the equations of motion for physical time evolution of the homogeneous YM condensate in the cosmological  environment. For this purpose, we first consider the effective QCD theory in the one-loop approximation, as in Appendix~\ref{Sec:matter}. As was shown in Ref.~\cite{Addazi:2018fyo}, an extrapolation of the one-loop approximated Lagrangian of the SU(2) gauge theory into deeply infrared (strongly-coupled) regime for the YM vacuum is numerically justified by a comparison to the all-loop result. This argument enables us to pursue the same path for QCD revealing the most important features of the generic non-perturbative QCD. Besides, the results of such a model can be useful to describe the properties of the homogeneous YM condensates at large $\mathcal{J}$, away from the ground state relevant for cosmological QCD phase transition.

\subsection{Einstein-YM equations}
\label{Sec:EYM}
\noindent 
We take a simplistic approach assuming that before the QCD transition epoch the gluon-field energy density is dominated by positively-valued chromomagnetic components, while negatively-valued chromoelectric components are negligibly small. For convenience, in what follows we re-label the corresponding energy-momentum tensor of the homogeneous YM condensate as $T^{\nu}_{\mu} \to T^{\nu,{\rm U}}_{\mu}$. As discussed in Refs.~\cite{Addazi:2018fyo}, both components approach asymptotic ground-state attractors, with exactly opposite densities, thus yielding the exact cancellation in the IR limit of the theory. In this section, we look at the dominant chromomagnetic solution in the quark-gluon plasma phase, and study its real time evolution relevant for the QCD transition epoch, away from the asymptotic ground-state attractor.

By the variational principle, one obtains the EYM system of operator equations of motion in a non-trivial spacetime
\begin{eqnarray}
&&\frac{1}{\varkappa}\left(R_\mu^\nu-\frac12\delta_\mu^\nu R\right)=
\frac{b}{32\pi^2}\frac{1}{\sqrt{-g}}\biggl[\biggl(-\mathcal{F}_{\mu\lambda}^a\mathcal{F}^{\nu\lambda}_a
\nonumber
\\
&& \quad +\,\frac14\delta_\mu^\nu
\mathcal{F}_{\sigma\lambda}^a\mathcal{F}^{\sigma\lambda}_a\biggr)
\ln\frac{e|\mathcal{F}_{\alpha\beta}^a\mathcal{F}^{\alpha\beta}_a|}{\sqrt{-g}\,
\lambda^4}-\frac14 \delta_\mu^\nu \,
\mathcal{F}_{\sigma\lambda}^a\mathcal{F}^{\sigma\lambda}_a\biggr]\,,\label{maineq} \\
&& \left(\frac{\delta^{ab}}{\sqrt{-g}}\partial_\nu\sqrt{-g}-f^{abc}\mathcal{A}_\nu^c\right)
\left(\frac{\mathcal{F}_b^{\mu\nu}}{\sqrt{-g}}\,\ln\frac{e|\mathcal{F}_{\alpha\beta}^a
\mathcal{F}^{\alpha\beta}_a|}{\sqrt{-g}\,\lambda^4}\right)=0\,. \nonumber
\end{eqnarray}
Notice that from now on, in all the derivations we will perform a rescaling of the gluon condensate, namely $\bar{g}\, U(t) \to U(t)$. 
The energy density and the pressure of other forms of matter are irrelevant for the discussion of the dynamical properties of the gluon condensate in the early Universe (in particular, for cosmological inflation), and will thus be omitted in practical calculations.

In what follows, we work in the flat FLRW conformal metric, characterized by the relations 
\[
\sqrt{-g}=a^4(\eta)\,, \qquad t = \int a(\eta) d\eta \,,
\]
the comoving time being defined in $ds^2=dt^2-a^2(t)\vec{dx}^2$. Besides this, 
we neglect quantum-wave fluctuations $\widetilde{S}_{ak}$, assuming 
that the homogeneous gluon condensate $U(t)$ dominates at considered spacetime scales. Under these conditions, the system of equations of motion describing conformal time evolution of the gluon condensate $U=U(\eta)$ and the scale factor $a=a(\eta)$ read
\begin{eqnarray}
&& \frac{6}{\varkappa} \frac{a''}{a^3} = T^{\mu,{\rm U}}_{\mu}\,, \nonumber \\
&& T^{\mu,{\rm U}}_{\mu}=\frac{3b}{16\pi^2 a^4}\Big[(U')^2-\frac{1}{4}U^4\Big]\,, \nonumber \\
&& \frac{\partial}{\partial \eta}\Big(U'\,\ln\frac{6e\big|(U')^2-\frac{1}{4}U^4\big|}{a^4\lambda^4}\Big) \nonumber \\ 
&& \qquad +\,\frac{1}{2}U^3\,\ln\frac{6e\big|(U')^2-\frac{1}{4}U^4\big|}{a^4\lambda^4}=0 \,. \label{eqU}
\end{eqnarray}
An additional coefficient $1/2$ appears in front of the QCD coupling constant --- which has been absorbed into the definition of the gluon field --- as compared to the SU(2) condensate case considered in Ref.~\cite{Pasechnik:2013sga}. The first integral of Eq.~(\ref{eqU}) is the Einstein $(0,0)$-equation, and reads
\begin{eqnarray}
&& \frac{3}{\varkappa}\frac{(a')^2}{a^4}=T^{0,{\rm U}}_{0}\,, \nonumber \\
&& T^{0,{\rm U}}_{0}=\frac{3b}{64\pi^2 a^4}\,\Big(\Big[(U')^2+\frac{1}{4}U^4\Big]\,
\ln\frac{6e\big|(U')^2-\frac{1}{4}U^4\big|}{a^4\lambda^4} \nonumber \\
&& \qquad +\, (U')^2 - \frac{1}{4}U^4\Big)\,. \label{eqUint}
\end{eqnarray}

\subsection{Asymptotic attractor solution and Z2 symmetry restoration}
\label{Sec:Z2}
\noindent 
In full analogy to the SU(2) condensate case considered in Ref.~\cite{Pasechnik:2013sga}, in the QCD case the system of equations (\ref{eqU}) and (\ref{eqUint}) has the exact solution corresponding to the vanishing logarithm or, equivalently, satisfies the transcendent equation
\begin{eqnarray}
|Q|=1\,, \qquad Q &\equiv& \frac{32 \pi^2 e}{11(\xi \Lambda_{\rm QCD})^4}T^{\mu,{\rm U}*}_{\mu} \nonumber \\ 
&=& \frac{6e\big[(U')^2-\frac{1}{4}U^4\big]}{a^4(\xi \Lambda_{\rm QCD})^4}\,, \label{Uexact}
\end{eqnarray}
which yields two distinct cases $Q=\pm 1$. 

One of these cases corresponding to $Q=+1$ has been discussed in the broad literature before (see e.g. Refs.~\cite{Zhang:1994pm,Xia:2007eu,Wang:2008fx,Dona:2015xia,Zhang:2002tt,Pasechnik:2013sga}). In particular, it provides a positive constant energy density of the gluon condensate
\begin{eqnarray}
(U')^2 \!-\! \frac{1}{4}U^4>0\,, \;\; T^{0,{\rm U}*}_{0}\equiv \frac{3b}{64\pi^2}\frac{(\xi \Lambda_{\rm QCD})^4}{6e} > 0 , \label{YM-dens-an}
\end{eqnarray}
and thus has been considered as a potential driver of early-time acceleration epochs in cosmology.

The second solution, characterized by $Q=-1$, has not been sufficiently discussed in the literature. It is worth noticing that it appears due to the symmetry of the corresponding RG equation with respect to the $\mathcal{J} \leftrightarrow -\mathcal{J}$ interchange --- see e.g. Ref.~\cite{Addazi:2018fyo}. Thus, this must be considered on the same footing as the solution with $Q=+1$. Such a contribution to the energy density of the universe is negative and has the same absolute value as in Eq.~(\ref{YM-dens-an}), i.e.
\begin{eqnarray}
\label{Qm1}
\!\!\!\!(U')^2 \!-\! \frac{1}{4}U^4<0, \;\; T^{0,{\rm U}*}_{0}\equiv \!-\frac{3b}{64\pi^2}\,\frac{(\xi \Lambda_{\rm QCD})^4}{6e} < 0 .
\end{eqnarray}
Provided such a negative-energy vacuum solution is stable (see Ref.), a positive CC-term should be present to compensate this negative contribution in order to comply with cosmological observations. One may notice, however, that the corresponding negative energy density for $\xi\simeq 4$ coincides with the quantum-topological term $\epsilon^{\rm QCD}_{\rm top.}$ provided in Eq.~(\ref{eps}). As was advocated in Ref.~\cite{Addazi:2018fyo}, such a cancellation is due to the emergence of a $\mathbb{Z}_2$ symmetry in the asymptotic ground state. The metastable positive-energy vacuum configurations are trapped in pockets of 3-space, which are separated from the stable CM regions by means of the infinite barriers, the so called domain walls.

Such a compensation mechanism grossly reduces or eliminates the QCD vacuum effect on the macroscopic late-time Universe expansion. 
Indeed, under the conditions specified in Ref.~\cite{Addazi:2018fyo}, the macroscopic evolution of the Universe reduces to the standard Friedmann equation, driven only by matter fields and a small uncompensated observable term $\epsilon_{\rm CC}\lll T^{0,{\rm U}*}_{0}$, while the evolution of the gluon condensate happens at characteristic microscopic scales corresponding to the QCD confinement scale $\Lambda_{\rm QCD}$, i.e.
\begin{eqnarray}
&& \frac{3}{\varkappa}\frac{(a')^2}{a^4}=\epsilon + \epsilon_{\rm CC}\,,\nonumber \\
&& (U')^2-\frac{1}{4}U^4=a^4 \frac{(\xi\Lambda_{\rm QCD})^4}{6e}\,, \qquad \xi\simeq 4 \,. \label{Uexact1}
\end{eqnarray}
Consequently, such a relatively slow macroscopic evolution of the Universe for $a=a(\eta)$, and the rapid fluctuations of the gluon condensate $U=U(\eta)$ at the characteristic QCD timescale, get practically separated and are independent from each other.

In the present Universe, for which we pick $a\equiv a_0 =1$, the exact (implicit) partial solutions for the homogeneous gluon condensate read
\begin{eqnarray}
&& Q=\pm 1\,, \qquad \int_{\widetilde{U}_0}^{\widetilde{U}}\frac{du}{\sqrt{\frac{1}{4}u^4 \pm 1}}=\widetilde{\eta}\,, \nonumber \\
&& \widetilde{U}=U\frac{(6e)^{1/4}}{4\Lambda_{\rm QCD}}\,,\qquad \widetilde{\eta}=
\eta\frac{4\Lambda_{\rm QCD}}{(6e)^{1/4}}\,, \label{Uexact2}
\end{eqnarray}
corresponding to (\ref{YM-dens-an}) and (\ref{Qm1}) solutions. 
Thus, the cosmological evolution of the gluon field in its ground state can be interpreted as a regular sequence of quantum tunneling transitions through the ``time-barriers'' represented by the regular singularities in the quantum vacuum solution of the effective YM theory. In this sense, the homogeneous gluon condensate in Minkowski spacetime is analogous to the topological condensate in the instanton theory of the QCD vacuum in Euclidean spacetime. This latter can be interpreted in terms of spatially-inhomogeneous gluon field fluctuations, which are induced by the quantum tunneling of the field through topological (spatial) barriers between different classical vacua.

It is worth noticing that the well-known 't Hooft-Polyakov monopole \cite{tHooft,Polyakov} is analogous to the classical YM condensate (\ref{SSU2}). While the monopole is introduced by means of an antisymmetric matrix with mixed Lorentz-isotopic indices in the Weyl gauge, the considered YM condensate provides a symmetric analogue of this solution.The singularities emerge at the level of the 4-potential of the YM condensate $U(t)$, while all the observable quantities (such as pressure and energy density), are finite. This situation reminds the Dirac monopole solution, where a singularity in the 4-potential emerges along the Dirac string, while the magnetic flux is finite.

In principle, the compensation may not be exact at the early stages of the Universe evolution, and may be fulfilled only at asymptotically large times and on the average. Notice furthermore that the macroscopic evolution of the present Universe, in this case, is only affected by a small remnant of the above cancellation, the observable cosmological constant term. The latter can be generated e.g. by an uncompensated quantum-gravity correction to the ground state energy in QCD, which may explain both its absolute value and sign, as was argued in Ref.~\cite{Pasechnik:2013poa}. Besides that, a positive uncompensated component of the YM vacuum, corresponding to the positive-valued homogeneous YM (non-QCD) condensate contribution in the early Universe, can be a natural cause for the cosmic inflation. Below, we will briefly discuss a possible scenario of the early-time accelerated expansion naturally provided by the homogeneous YM condensate $U(t)$.

\subsection{Numerical analysis and tracker solutions of the EYM system}
\label{Sec:num}
\noindent 
We now consider a deviation from the exact partial solution given by Eq.~(\ref{Uexact2}), and study the general solution of the equations of motion (\ref{eqU}) and (\ref{eqUint}), numerically. Let us first choose the subset of initial conditions satisfying $Q_0\equiv Q(t=t_0)>1$, and discuss the results of the numerical analysis qualitatively.
\begin{figure}[!h]
\begin{minipage}{0.32\textwidth}
 \centerline{\includegraphics[width=1.27\textwidth]{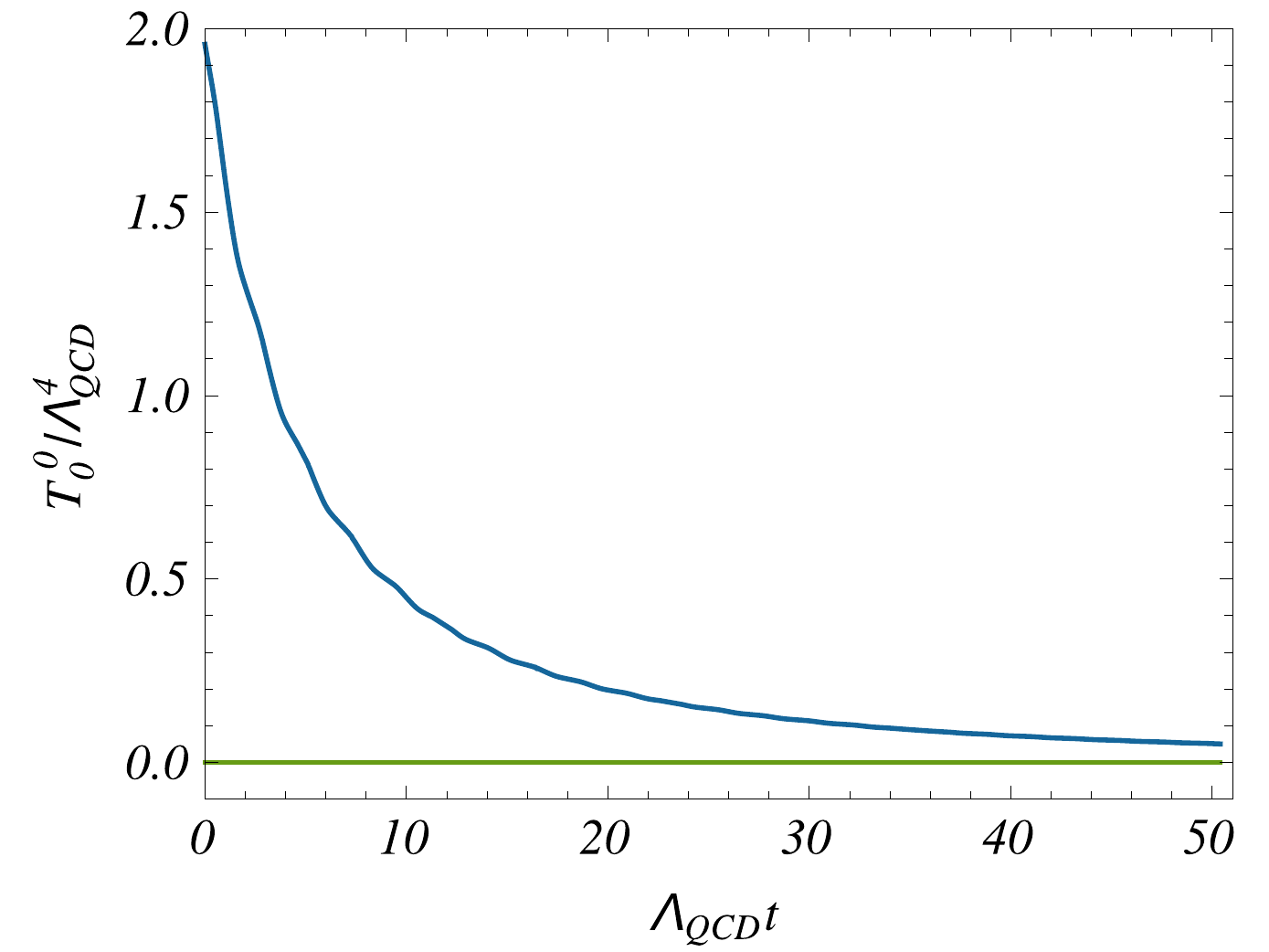}}
\end{minipage}
\begin{minipage}{0.32\textwidth}
 \centerline{\includegraphics[width=1.27\textwidth]{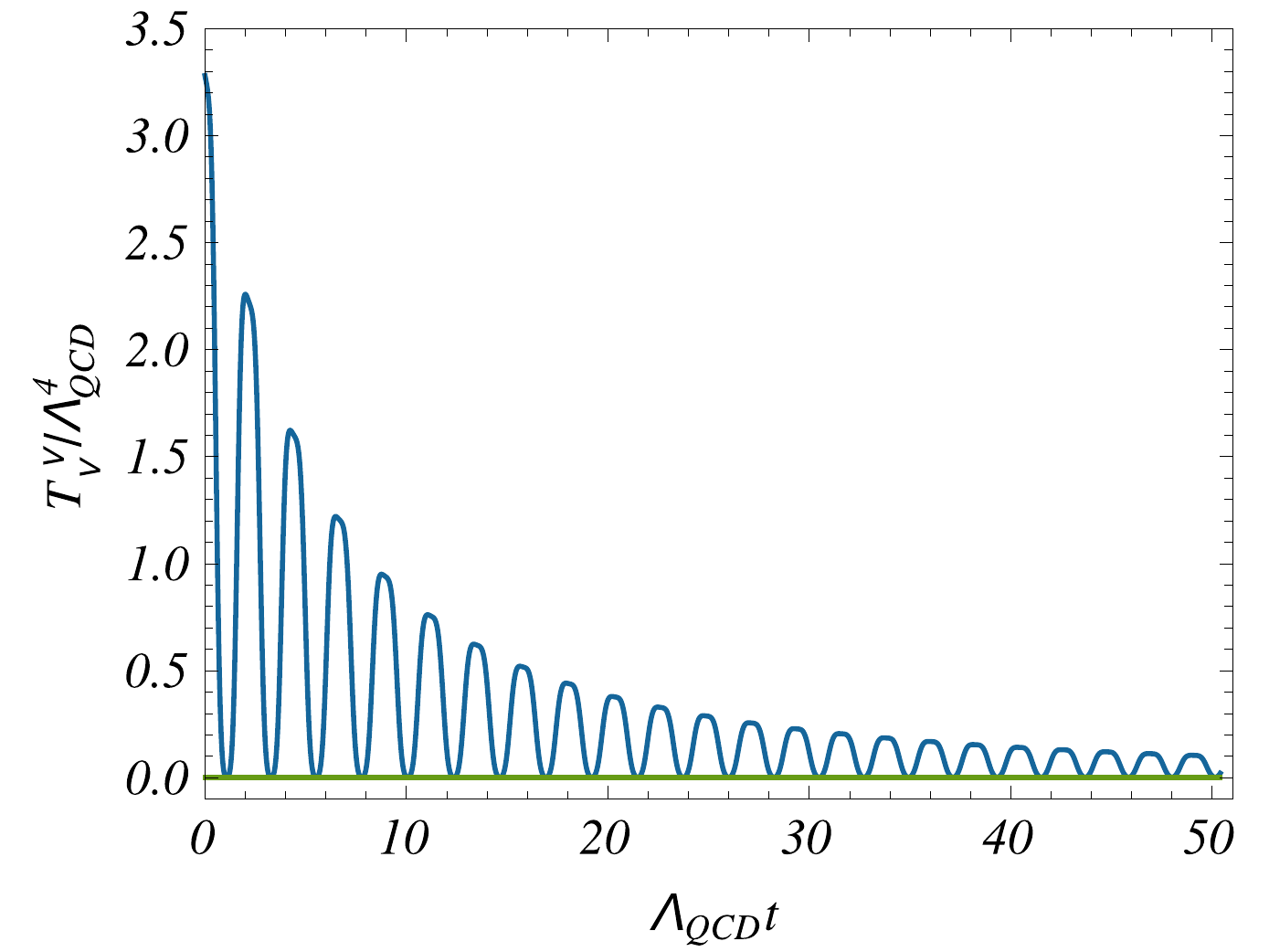}}
\end{minipage}
\begin{minipage}{0.32\textwidth}
\centerline{\includegraphics[width=1.25\textwidth]{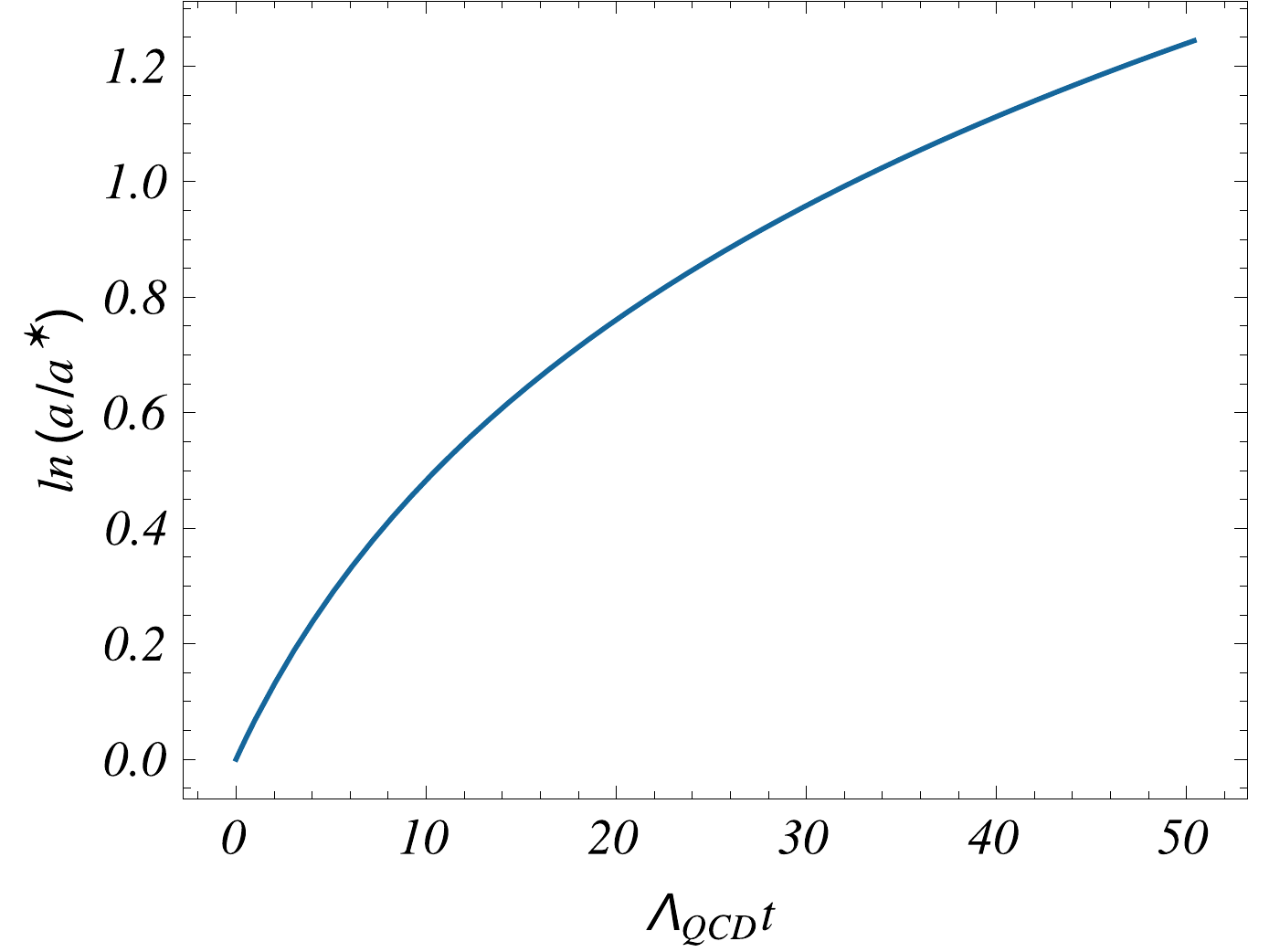}}
\end{minipage}
\caption{We illustrate, as functions of the physical time $t=\int ad\eta$, and using units of the characteristic time scale $\Lambda_{\rm QCD}^{-1}$, respectively the total energy density $T^0_0(t)$ of the homogeneous gluon condensate (upper panel), the trace of the total QCD energy-momentum tensor $T^\mu_\mu(t)$ (middle panel) and the logarithm of the scale factor $a(t)$ (lower panel). We indicate the total energy density and the trace values for $Q_0\equiv Q(t_0)=1$, respectively, with horizontal lines in the up and middle panels, where the initial conditions are chosen as $U_0=0$, $\dot{U}_0=(\xi \Lambda_{\rm QCD})^2/\sqrt{3e}$, $Q_0>1$, $\xi\simeq 4$, and 
the gravitational constant is set to be $\varkappa=10^{-7} {\rm MeV}^{-2}$, for simplicity of the numerical analysis. We plot both $T^0_0(t)$ and $T^\mu_\mu(t)$ in dimensionless units, and rescale them by $\Lambda_{\rm QCD}^4$. It is evident that the amplitude of the quasi-periodic oscillations of $Q=Q(t)$ happens to decrease at large $t\gg \Lambda_{\rm QCD}^{-1}$, and to approach asymptotically unity. 
}
\label{fig:YMC-1}
\end{figure}
For this choice of initial conditions, Fig.~\ref{fig:YMC-1} (upper panel) illustrates the physical time evolution of the total energy density (in dimensionless units) of the homogeneous gluon condensate $U=U(t)$
\begin{eqnarray}
\label{QCD-vac}
T^0_0(t) \equiv \bar{\epsilon} + T^{0,{\rm U}}_0(t) \,, 
\end{eqnarray}
where $T^{0,{\rm U}}_0$ and $\bar{\epsilon}$ are given respectively by Eqs.~(\ref{eqUint}) and by $\bar{\epsilon}=\epsilon_{\rm top.}^{\rm (QCD)}+\epsilon_{\rm CC}$, the $\epsilon_{\rm CC}$ term denoting the cosmological constant contribution. In Fig.~\ref{fig:YMC-1} (middle panel), we show the corresponding result for the trace of the total gluon energy-momentum tensor 
\begin{equation}
\label{QCDtrace}
T^\mu_\mu(t)\equiv 4\bar{\epsilon} + T^{\mu,{\rm U}}_\mu(t) \,,
\end{equation}
in dimensionless units, and the corresponding solution for the logarithm of the scale factor is given in Fig.~\ref{fig:YMC-1} (lower panel). The period of the $T^\mu_\mu(t)$ oscillations is practically time independent, which can also be proven analytically (see below), while a small residual time-dependence appears due to a possibly large deviation from $Q=1$. In this analysis we used $\xi\simeq 4$, providing the exact compensation of the QCD vacuum 
energy density at $t\gg t_0$, such that a change of $\xi$ will only affect the asymptotic values of $T^0_0(t)$ and $T^\mu_\mu(t)$ at large $t$.

\begin{figure}[!h]
\begin{minipage}{0.4\textwidth}
 \centerline{\includegraphics[width=1.0\textwidth]{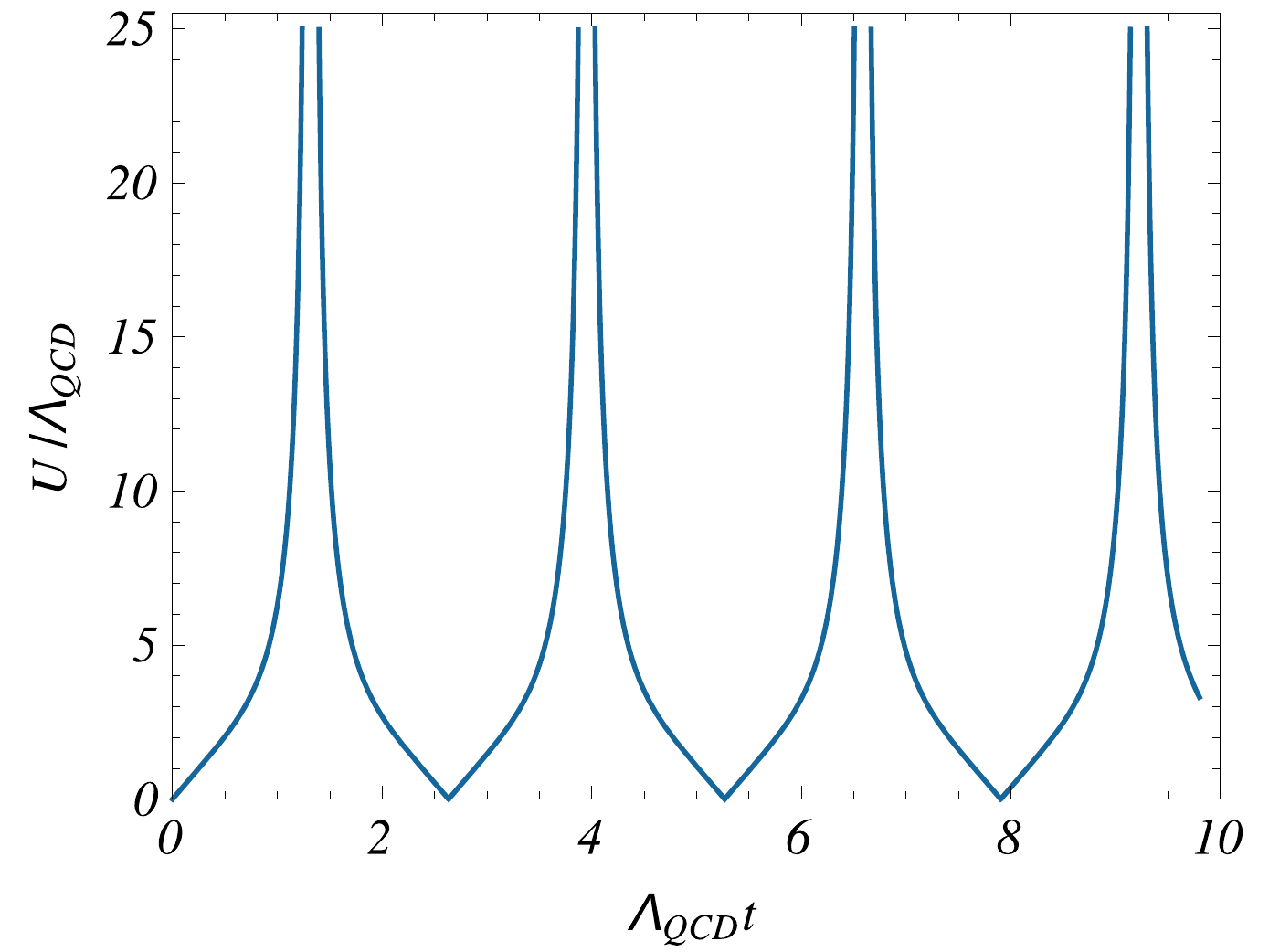}}
\end{minipage}
\hspace{1cm}
\begin{minipage}{0.4\textwidth}
 \centerline{\includegraphics[width=1.0\textwidth]{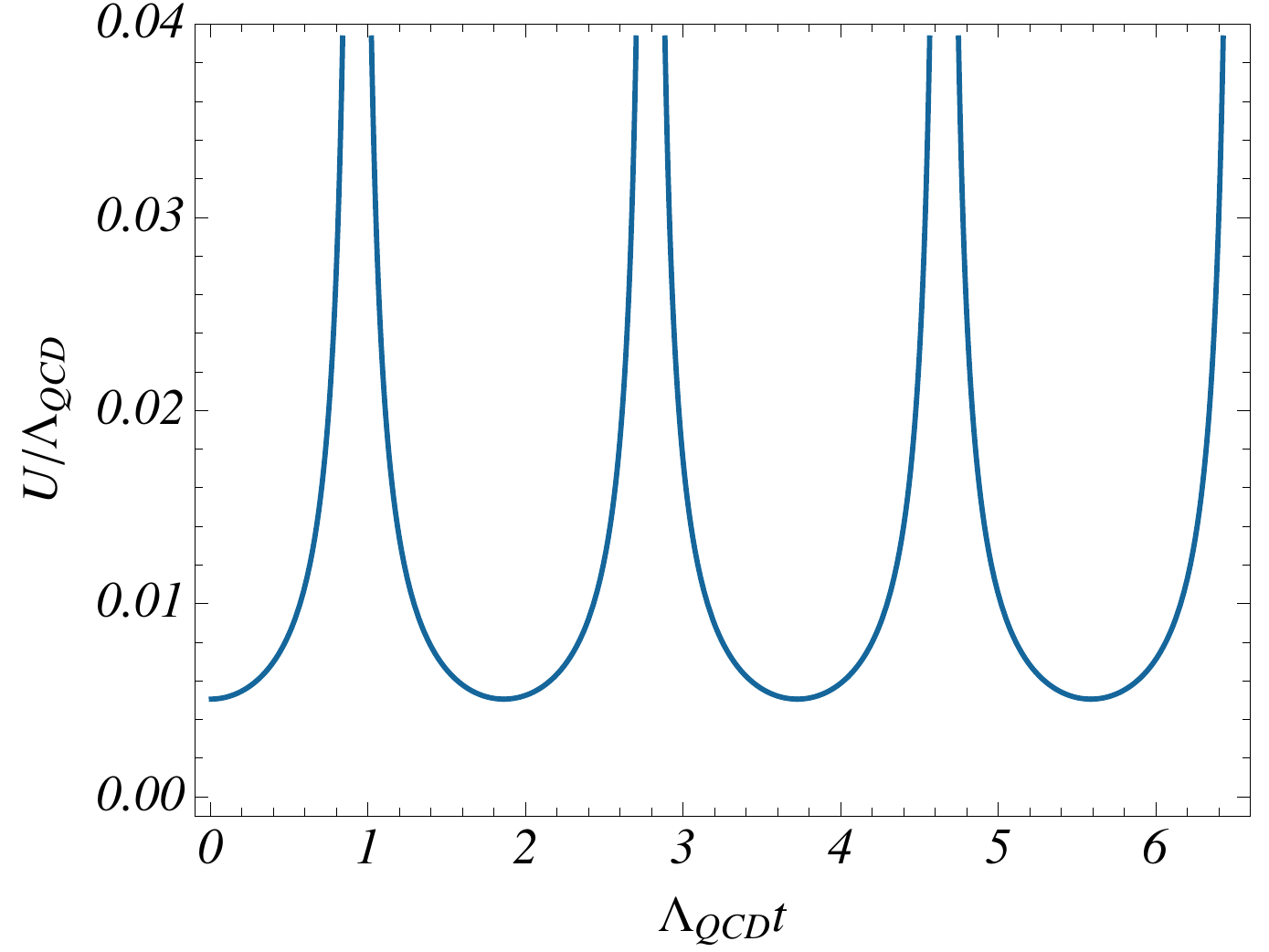}}
\end{minipage}
\caption{We show the homogeneous QCD condensate amplitude oscillations $U=U(t)$, manifesting quasi-periodic singularities in the physical time $t=\int ad\eta$, for the solutions labelled by $Q(U)=1$ and the $Q(U)=-1$. These are shown respectively in the upper and lower panels, using units of the characteristic time scale $\Lambda_{\rm QCD}^{-1}$. The spikes that are displayed are localized in time-lapse, along the space-like directions, and must be interpreted as new solitonic solutions. In a previous work, we dubbed these solutions {\it chronons}, or {\it $\chi$-solutions}.}
\label{fig:Uexact}
\end{figure}

Although the amplitude of the condensate $U(t)$ possesses quasi-periodic singularities, as is seen in the upper panel of Fig.~\ref{fig:Uexact}, the evolution of its energy density $T^{0,{\rm U}}_0(t)$ as well as pressure (or $T^{\mu,{\rm U}}_\mu(t)$) remains 
continuous in time. One immediately notices that the general solution asymptotically reaches the analytic solution 
corresponding to Eq.~(\ref{YM-dens-an}). This happens after a number of oscillations of function $Q(t)$ whose amplitude 
approaches unity at large physical times $t$, $Q(t\to \infty)\Rightarrow 1$, for any initial conditions satisfying $Q_0>0$. 
During such a relaxation regime, the total energy density of the QCD vacuum continuously decreases and eventually vanishes 
in the asymptotic limit $t\gg t_0$. Note, this regime is accompanied by a decelerating expansion of the Universe.

The same quantities have also been studied in the opposite case of initial conditions, i.e.  for $0<Q_0<1$, and the results (without accounting for a $\bar{\epsilon}$ term) are illustrated in the lower panel of Fig.~\ref{fig:Uexact}. In this case the general solution asymptotically approaches the de-Sitter regime as well, in full analogy with the $Q_0>1$ case. A qualitatively similar situation is realized for $Q_0<0$ as well. The de-Sitter solution, therefore, appears as an attractor (or tracker) solution of the EYM system. This provides a dynamical mechanism for the elimination of the gluon vacuum component of the ground state energy of the Universe, asymptotically at macroscopic space-time scales and for arbitrary initial conditions and parameters of the model. This fact provides a generalization of the basic result of Ref.~\cite{Pasechnik:2013sga} to the case of arbitrary initial conditions and to a gauge group possessing a $SU(2)$ subgroup.

\section{Asymptotic behaviour of the gluon condensate}
\label{Sec:asympt}
\noindent 
In Section~\ref{Sec:num}, the system of equations (\ref{eqU}) and (\ref{eqUint}) was investigated numerically in the general case. We assumed arbitrary initial conditions and found the universal asymptotics corresponding to the (partial) exact analytic solutions with minimal energy $|Q|=1$. In this Section, we will construct the general analytic approximate solution of this system, in the case of $Q(t)$ being not too far from unity which reproduces the results of numerical simulations and provides us with an extra important information about the cosmological evolution of the QCD (or YM, in general) vacuum.  In particular, such a solution will provide an additional proof for the observed asymptotic behaviour of $a(t)$, $T_0^0(t)$ and $T_{\mu}^{\mu}(t)$. By comparison with the numerical analysis performed above, we will notice that all the characteristic features of the approximate analytic solution, corresponding to the one-loop effective action, turn out to be valid also in the case of a large deviation of $Q(t)$ from unity, as well as beyond the perturbation theory. In this sense, the aforementioned numerical analysis is of important guidance, and supplements the approximated analytic results obtained below.

\subsection{Period of oscillations and relaxation time}
\label{Sec:T}
\noindent 
One could expect that the oscillatory behavior of the $U(\eta)$ function found analytically for $|Q|=1$, will also be present for $Q_0\neq 1$, as long as the deviation of $Q_0$ from unity is not very large. Indeed, if the initial energy density is not too large than the period of oscillations of the positive-energy solution for the condensate $U=U(\eta)$ in conformal time, $T_{\eta}^{>}$ can be roughly estimated using Eq.~(\ref{Uexact2}) as follows
\begin{eqnarray} 
T_{\eta}^{>}\simeq \frac{4k\, (6e)^{1/4}}{a \xi \Lambda_{\rm QCD}}\,, \qquad 
k&\equiv& \int_0^{\infty} \frac{du}{\sqrt{1+\frac{1}{4} u^4}} \nonumber \\ 
&=& \frac{\Gamma(1/4)^2}{2\sqrt{2\pi}}\approx 2.622 \,. \label{Teta}
\end{eqnarray}
The corresponding period in physical time reads approximately
\begin{eqnarray}
\label{new2}
T_t \equiv T_t^{>}\simeq \frac{4k\, (6e)^{1/4}}{\xi \Lambda_{\rm QCD}} \simeq \frac{5.3}{\Lambda_{\rm QCD}}\,, 
\end{eqnarray}
which is close to the numerical result discussed above. Notice that the period of YM condensate oscillations corresponding to the negative-energy solution (\ref{Qm1}) $T_t^{<}$ differs from Eq.~(\ref{new2}), being
\begin{eqnarray}
\label{new2-1}
T_t^{<} \simeq \frac{\sqrt{2}k\, (6e)^{1/4}}{\xi \Lambda_{\rm QCD}} \simeq \frac{1.86}{\Lambda_{\rm QCD}}\,.
\end{eqnarray}
The conformal time derivative of the gluon condensate energy density for $Q_0 > 1$ in QCD, given by
\begin{eqnarray}
\!\!\!\frac{\partial T_0^{0,{\rm U}}}{\partial \eta} \!\!=\!\! - \frac{33 \varkappa}{16 \pi^2}\frac{a'}{a^5}\!\!\Big[U'^2\!+\!\frac{1}{4}U^4\Big] \!\!
\ln\!\frac{6e\big|(U')^2-\frac{1}{4}U^4\big|}{a^4(\xi \Lambda_{\rm QCD})^4} ,
\end{eqnarray}
is negative in the initial moment of time $t=t_0$ and its absolute value decreases such that $T_0^{0,{\rm U}}(t)$ indeed approaches the value $T_0^{0,{\rm U}^*}$, corresponding to the exact analytical (de-Sitter) solution given by Eq.~(\ref{YM-dens-an}). 
On dimensional grounds, a rough estimate for the relaxation time of the YM condensate energy density is approximately given by
\begin{eqnarray}
\label{new3}
t_r\simeq \frac{1}{\sqrt{\varkappa \epsilon_0}}\,, \qquad \epsilon_0\equiv T_0^0(t=t_0) \,.
\end{eqnarray}
As will be shown analytically below, the scale factor approaches the de-Sitter solution at late times
\begin{equation}
t>\frac{1}{\sqrt{\varkappa \epsilon_{\rm CC}}} \gg t_r \,.
\end{equation}

Under the viable hypothesis of co-existence of chromoelectric and chromomagnetic condensates around the QCD phase transition epoch,their initial values at $t=t_0$ before the hadronization of the quark-gluon plasma may be different, and thus they may not identically compensate each other. Due to the attractor nature of the corresponding solutions the compensation may happen only at late times (in particular, after the Universe gets hadronized), at typical time scales $t\gg t_r$. While the compensation can be (nearly) exact when the ground-state energy density is averaged over macroscopically large spacetime separations, such a compensation may not be exact locally, yielding a possibly non-vanishing effect (within a typical hadron scale one is sensitive to) in e.g. hadronic reactions.

Now let us consider the oscillations of the trace $T_{\mu}^{\mu,{\rm U}}(t)$, providing an information about the pressure of the gluon condensate. The unobservable function $U(t)$, and hence the combination $(U')^2+\frac{1}{4}U^4$, oscillate with non-physical quasi-periodic singularities. Nonetheless, the energy density $T_0^{0,{\rm U}}(t)$, which is a physical observable given by Eq.~(\ref{eqUint}), has to be continuous, as confirmed by the numerical analysis. The latter condition can only be realized if $Q=Q(t)$ oscillates as well with the same period as the condensate,
and periodically reaches unity (such that its logarithm vanishes). In such a way $Q=Q(t)$ compensates the corresponding singularities in the coefficient $(U')^2+\frac{1}{4}U^4$ term. Besides that, the $Q=Q(t)$ function can not cross unity, since otherwise it would lead to a non-continuity in the behavior of $T_0^{0,{\rm U}}(t)$, due to a sign change in the first term in Ref.~(\ref{eqUint}) when $Q=1$. From these qualitative arguments it is clear that the time evolution of $Q=Q(t)$ function satisfies the constraints $0<Q(t)\leq1$ or $Q(t)\geq1$ for the initial conditions $0<Q_0<1$ or $Q_0>1$, respectively. The period of its oscillations can then be estimated as
\begin{eqnarray}
\label{gT}
T\equiv \frac{T_t}{2} \simeq \frac{2k\, (6e)^{1/4}}{\xi \Lambda_{\rm QCD}} \,,
\end{eqnarray}
and its relaxation time is the same as for $T_0^{0,{\rm U}}(t)$ given in Eq.~(\ref{new3}). Consequently, these basic features of the general solution of the EYM equations for the YM condensate (as the periods of oscillations and the relaxation times of the condensate), its energy density and the pressure can be described qualitatively, without a reference either to the numerical calculations or to a particular choice of model parameters.

\vspace*{3mm}

\acknowledgments 
\noindent 
A.A and A.M. acknowledge support by the NSFC, through the grant No. 11875113, the Shanghai Municipality, through the grant No. KBH1512299, and by Fudan University, through the grant No. JJH1512105. R.P. was partially supported by the Swedish Research Council, contract numbers 621-2013-428 and 2016-05996, by CONICYT grant MEC80170112 (Chile), as well as by the European Research Council (ERC) under
the European Union's Horizon 2020 research and innovation programme (grant agreement No 668679). This work was supported in part by the Ministry of Education, Youth and Sports of the Czech Republic, 
project LT17018. The work has been performed in the framework of COST Action CA15213 ``Theory of hot matter and relativistic heavy-ion collisions'' (THOR).

\end{document}